\documentclass[fleqn,twoside]{article}
\usepackage{latexsym}
\usepackage{espcrc2}
\usepackage[xdvi]{graphicx}

\newcommand{\hmu}{\hat{\mu}}
\newcommand{\der}{\partial}
\newcommand{\ddx}[1]{\frac{\partial}{\partial #1}}

\newcommand{\vol}{\mathit{vol}}
\newcommand{\zform}{{\mbox{\scriptsize $0$-form}}}

\newcommand{\tr}{\mathop{\mathrm{tr}}\nolimits}

\newcommand{\unity}{\mbox{\unboldmath$\lefteqn{\mathsf{1}}%
\hspace*{0.12em}\mathsf{1}$}}
\newcommand{\psibar}{\bar{\psi}}
\newcommand{\Psibar}{\bar{\Psi}}
\newcommand{\varphibar}{\bar{\varphi}}

\newcommand{\linkns}[4]{
 \setlength{\unitlength}{.1em}
 \begin{picture}(30,15)(8,5)
  \thicklines
  \put(10,5){\vector(1,0){20}}
  \put(30,5){\line(-2,1){5}}
  \put(30,5){\line(-2,-1){5}}
  \put(10,5){\makebox(0,0){#1}}
  \put(30,5){\makebox(0,0){#2}}
  \put(10,13){\makebox(0,0){#3}}
  \put(30,13){\makebox(0,0){#4}}
 \end{picture}
}

\title{Dirac-K\"ahler fermion with noncommutative differential forms on
       a lattice\thanks{Talk presented by I. Kanamori.}}
\author{I. Kanamori\address{Department of Physics, Hokkaido University,
        Sapporo, 060-0810, Japan}
        and 
        N. Kawamoto\addressmark
}

\begin{document}

\begin{abstract}
Noncommutativity between a differential form and a function allows us
to define differential operator satisfying Leibniz's rule on a lattice.
We propose a new associative Clifford product defined on 
the lattice by introducing the noncommutative differential forms.
We show that this Clifford product naturally leads to the Dirac-K\"ahler 
fermion on the lattice.
\end{abstract}

\maketitle

\section{Noncommutative differential form on a lattice}
\label{sec:ncdf}
Let us first point out that the lattice difference operator, defined by 
$  df(x) = \sum_\mu \der_{+\mu}f(x) dx^\mu $ with 
$\der_{+\mu}f(x) = f(x+\hmu) - f(x)$, does not satisfy the Leibniz's 
rule, where $x+\hmu$ denotes the next neighbor lattice site of $x$ into 
$\mu$ direction. Instead the following relation holds:  
\begin{eqnarray}
 d( f(x)g(x) )
  &=& (\der_{+\mu}f(x))g(x+\hmu)dx^\mu  \nonumber\\
  &&  {}+ f(x) \der_{+\mu}g(x) dx^\mu.
\end{eqnarray}
If we, however, introduce the following noncommutativity between functions 
and forms, 
$
g(x+\hmu) dx^\mu = dx^\mu g(x),
$
\cite{Dimakis:1992pk,Aschieri:1993wg}, 
we can define the differential operator satisfying the following lattice 
Leibniz's rule: 
\begin{equation}
 d(f(x)g(x)) =  (df(x))g(x) + f(x)dg(x).
\end{equation}

We may be able to understand the noncommutative nature between the 
functions and a differential as follows.
We identify that $0$-form lives on a site while $1$-form lives on a link.
A link has two boundary sites so that a function sitting on one site of the 
link and a function sitting on the other site may induce a difference in 
the order of the product. Pictorially we denote the situation as: 
\begin{eqnarray*}
  g\,dx^\mu \sim 
   \linkns{$\bullet$}{}{$x$}{},
 &&
   dx^\mu\,g \sim 
   \linkns{}{$\bullet$}{}{$x+\hmu$}\ \ \ ,
\end{eqnarray*}
which denotes the noncommutativity.

\section{Dirac-K\"ahler fermion and Clifford product}
We first briefly summarize the formulation of Dirac-K\"ahler 
fermion\cite{I-L,Kahler:1962,Graf:1978kr,Becher:1982ud}.
Dirac-K\"ahler fermion uses inhomogeneous differential forms to describe
fermions. We first note the following well known
relations on the flat spacetime: 
\begin{equation}
(d -\delta)^2= \partial^\mu \partial_\mu = (\gamma^\mu\partial_\mu)^2,
\end{equation}
where $\delta$ is the adjoint of the exterior derivative $d$ which 
suggests that $d-\delta$ plays a role of Dirac operator in the space 
of differential forms. 
We introduce the following Clifford product $\vee$ \cite{Kahler:1962}:
\begin{equation}
 d-\delta
   = \left( dx^\mu + \ddx{dx^\mu}\right)\der_\mu
   \equiv dx^\mu \vee \der_\mu = d\vee,
\end{equation}
which satisfies the Clifford algebra,
\begin{eqnarray}
 dx^\mu \vee dx^\nu + dx^\nu \vee dx^\mu = 2g^{\mu\nu}.
\end{eqnarray}
$dx^\mu$ plays a role of $\gamma$ matrix in the space of differential 
forms with Clifford product.

The Dirac operator $d \vee$ mixes degrees of differential forms.
We define a fermion field as a direct sum of all degrees of forms:
\begin{eqnarray}
 \Psi = \varphi + \varphi_\mu\,dx^\mu
          + \varphi_{\mu\nu}\,dx^\mu \wedge dx^\nu + \cdots.
\end{eqnarray}
$\Psibar$ is defined in a similar way.
The Clifford product between such general forms is defined 
as \cite{Kahler:1962}:
\begin{eqnarray}
 \lefteqn{
  \Psi \vee \Psi'
  }\nonumber\\
  &\!\!\!=\!\!\!& \sum_{p=0}^{D} \frac{1}{p!} (-1)^{p(p-1)/2}
    \left( \eta^p \ddx{dx^{\mu_1}}\cdots\ddx{dx^{\mu_p}}\Psi \right)
      \nonumber\\
  &&   \wedge \left( \ddx{dx_{\mu_1}}\cdots\ddx{dx_{\mu_p}}\Psi' \right),
\end{eqnarray}
where 
$\eta \mbox{($p$-form)} = (-1)^p \mbox{($p$-form)}$ is a sign
factor to accommodate associativity.

We can then define the free fermion action:
\begin{eqnarray}
 2^{-D/2} \int \Psibar \vee (d+m) \vee \Psi \Bigm|_\zform \vol \\
 =  \int \sum_{(j)} \psibar_{(j)}
       ( \gamma^\mu \partial_\mu + m) \psi^{(j)} \vol.
\end{eqnarray}
Here, $m$ is a $0$-form mass term.
The action is known to give Dirac fermions with $2^{D/2}$ ``flavors''.
Components of the spinors are related to antisymmetric tensors through
\begin{equation}
 \psi{}_\alpha{}^{(j)}
   = \bigl(
        \varphi\unity + \varphi_\mu \gamma^\mu
        +  \varphi_{\mu\nu}\gamma^{[\mu} \gamma^{\nu]} + \cdots \bigr)
         {}_\alpha{}^{(j)},
\end{equation}
where $(j)$ can be generally recognized as a flavor suffix.

\section{Clifford product on the lattice}

Using the noncommutative differential forms on the lattice, we propose 
a new definition of Clifford product. A similar but non-associative 
result is found in ref.~\cite{Vaz1}.

We first point out that hermiticity of the action requires to introduce 
the symmetric lattice which has positively and negatively oriented lattice 
as shown in Fig. 1. 
Using the forward and backward difference operators: 
$\der_{\pm\mu}f(x) = \pm (f(x\pm\hmu) -f(x))$, we define the following 
exterior derivative operator on the symmetric lattice:
\begin{eqnarray}
  df = \sum_\mu \left\{(\der_{+\mu} f) \theta^{+\mu}
                       - (\der_{-\mu}f) \theta^{-\mu}\right\},
\end{eqnarray}
where $\theta^{+\mu}$ denotes $1$-form on the positively oriented links and
$\theta^{-\mu}$ denotes the negative counterpart.
\begin{figure}
 \begin{minipage}[c]{.35\linewidth}
  \includegraphics[width=\linewidth]{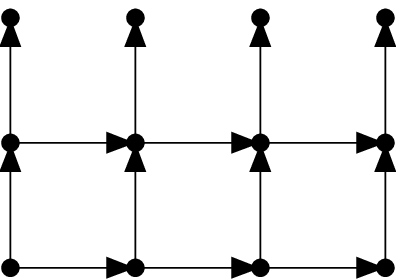}
 \end{minipage} 
 \begin{minipage}[c]{.27\linewidth}
  \setlength{\unitlength}{.01\linewidth}
  \begin{picture}(100,10)(0,0)
   \put(0,0){\vector(1,0){100}}
  \end{picture}
  \begin{center}
   Hermiticity
  \end{center} 
 \end{minipage}
 \begin{minipage}[c]{.35\linewidth}
  \includegraphics[width=\linewidth]{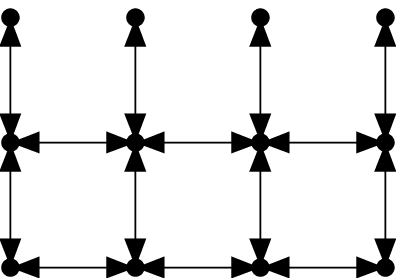}
 \end{minipage}
 \vspace*{-1em}

 \caption{A lattice with only positively oriented links(left) and both
 oriented links(right)}
 \label{fig:sym-lat}
\end{figure}
We introduce the following noncommutativity on the lattice:
\begin{equation}
 \theta^{\pm\mu}f(x) = f(x \pm \hmu) \theta^{\pm\mu}.
\end{equation}

We introduce a natural complex structure by $*-$conjugation, 
$(\theta^{\pm\mu})^* = -\theta^{\mp\mu}$.
The conjugation flips the orientation of the link, which reminds us 
of the relation of the gauge link variables $U_\mu$ and $U_\mu^\dagger$.
This type of noncommutative differential forms were used by several
authors\cite{
Dimakis:1994qq,Dai,Aschieri:2002vn}.

In order to define the Clifford product on the lattice, we require the 
following criteria: 1) associativity of the product, 2) positivity of an 
inner product defined by the Clifford product. There are, however, some 
arbitrariness for the definition of the Clifford product on the symmetric 
lattice since we have two types of one form; $\theta^{+\mu}$ and 
$\theta^{-\mu}$. 

After systematic investigations, we obtained the following 
definition of the Clifford product compatible with 
the above criteria\cite{Kanamori:2003mc}:
\begin{eqnarray}
 f \theta^K \vee g \theta^L
  &=& \sum_{p=0}^{2D}(-1)^{p(p-1)/2} \left(-\frac{1}{2}\right)^p
       \sum_{\{\epsilon_i \mu_i\}}
	\Bigl\{  \nonumber \\
  && \eta^p \Bigl(\prod_i \ddx{\theta^{\epsilon_i \mu_i}}
                             T_{\epsilon_i \mu_i}\Bigr)
                f \theta^K \Bigr\}
      \nonumber \\
  && {}
 \wedge \Bigl\{ \prod_i \ddx{\theta^{-\epsilon_i \mu_i}}
                       g \theta^L \Bigr\},
\end{eqnarray}
where we should not sum up permmutably equivalent product of the
differentiation of $\theta^{\pm\mu}$ in the summation.
$\theta^K$ is a short hand notation of $k$-form
$\theta^{\epsilon_1\mu_1} \ldots \theta^{\epsilon_k\mu_k}$, and 
$\theta^L$ is a similar $l$-form.   
$\epsilon_i\mu$ denotes $+\mu$ or $-\mu$.
The shift operator $T_{\epsilon\mu}$ generate a shift of argument,
$T_{\epsilon\mu}f(x) = f(x+\epsilon\hmu)$. Without the shift operators
we would not have obtained the associative Clifford product.

\section{Fermion with the lattice Clifford product}

Having defined the Clifford product on the lattice, we can formulate
Dirac-K\"ahler fermion on the lattice.
We choose to define the fermion and anti-fermion fields by the following 
differentials:
\begin{eqnarray}
 \Psi
   &=& \varphi + \varphi_\mu\,\sqrt{2}\theta^{+\mu} \nonumber \\
   &&  {}+ \varphi_{\mu\nu}\,(\sqrt{2}\theta^{+\mu})(\sqrt{2}\theta^{+\nu})
       + \cdots \\
 \Psibar
   &=& \varphibar -\sqrt{2}\theta^{-\mu}\,\varphibar_\mu \nonumber \\
   &&   {}+ (-\sqrt{2}\theta^{-\mu})(-\sqrt{2}\theta^{-\nu})\,
             \varphibar_{\mu\nu} + \cdots 
\end{eqnarray}
where $\Psi$ lives on the positively oriented lattice and $\Psibar$
lives on the negatively oriented lattice and they are on top of each
other.
The action is almost the same as the continuum one:
\begin{eqnarray}
 \lefteqn{
   \int \Psibar \vee ( \sqrt{2} d + m )
   \vee \Psi \Bigm|_\zform \vol
  } \nonumber \\
  &\!\!\!=\!\!\!&\! 2^{\frac{D}{2}}\sum_x \Bigl[\ 
     \frac{1}{2}\sum_\mu \Bigl\{
  \psibar_{(j)}(x) \gamma_\mu
         \bigl( \der_{+\mu} + \der_{-\mu}\bigr)\psi(x)^{(j)}
     \nonumber \\*
  && 
      \!\!\!{}+\psibar_{(j)}(x) \gamma_5^\dagger{}
         \bigl( \der_{+\mu} -\der_{-\mu} \bigr)\psi(x)^{(l)}
  (\gamma_5 \gamma_\mu)_{(l)}{}^{\!(j)}
     \Bigr\}
     \nonumber\\
  &\!\!&  \hspace{4em}
     {}+ m \psibar_{(j)}(x)\, \psi^{(j)}(x)
   \Bigr]
\label{eq:action}
\end{eqnarray}
The last expression was shown to be equivalent to the staggered lattice
fermion\cite{Gliozzi:1982ib,Kawamoto:1981hw}. 

Although staggered fermion is known as a latticized Dirac-K\"ahler
fermion in a flat spacetime, 
this kind of straightforward derivation was not known.

\section{Discussion}

We need to point out that the introduction of the gauge field 
can be straightforwardly accommodated by the use of the Clifford 
product which includes the known formulation as a particular 
case\cite{Dimakis:1992pk,Dimakis:1994qq}. 

The Yang-Mills action can be given by the following form by 
the use of Clifford product:
\begin{equation}
   S = -\frac{1}{2g^2}\tr\int F \vee F \Bigm|_\zform \vol
\end{equation}
where $F = dA + A^2$ is $2$-form curvature.
This formulation works both in the continuum and lattice spacetime.
In the lattice case, we obtain plaquette action as in \cite{Dimakis:1994qq}.
Thus we can treat boson and fermion by the same Clifford product 
with noncommutative differential form.

The Dirac-K\"ahler fermion on the lattice itself is an interesting
problem\cite{deBeauce:2003zd}. 
Here we point out two other possible applications.
In the supersymmetric field theory, Leibniz's rule plays an important
role
which was one of the difficulties of formulating 
SUSY on lattice.
We, however, have defined lattice derivatives satisfying the Leibniz's 
rule on the lattice by introducing noncommutativity. 
One of the authors pointed out that Dirac-K\"ahler fermion formalism is 
essentially equivalent to the twisting of topological field theory 
generating SUSY\cite{Kawamoto:1999zn,K-K-U}. 
We would like to show that present formalism with the Clifford product 
plays an important role in defining SUSY on the lattice.

Next application is to define gravity with matter fermion on the 
dynamically triangulated lattice. 
Generalization to the curved space time is expected to be straightforward 
since Dirac-K\"ahler fermion can be formulated by the differential forms 
with noncommutative differential forms and is independent on the lattice 
structure and thus can be defined on the triangulated lattice.

\end{document}